\def\real{{\tt I\kern-.2em{R}}}
\def\nat{{\tt I\kern-.2em{N}}}
\def\realp#1{{\tt I\kern-.2em{R}}^#1}
\def\natp#1{{\tt I\kern-.2em{N}}^#1}
\def\hyper#1{\ ^*\kern-.2em{#1}}

\def\st#1{{\tt st}(#1)}

\def\hyperrealp#1{{\tt ^*{I\kern-.2em{R}}}^#1} 
\def\hypernat{{^*{\nat }}}
\def\hypernatp#1{{{^*{{\tt I\kern-.2em{N}}}}}^#1} 
\def\eskip{\hskip.25em\relax}

\def\Hyper#1{\hyper {\eskip #1}}
\def\leaderfill{\leaders\hbox to 1em{\hss.\hss}\hfill}
\def\srealp#1{{\rm I\kern-.2em{R}}^#1}

\def\power#1{{{\cal P}(#1)}}

\def\pars{\par\smallskip}
\def\parm{\par\medskip}
\def\r#1{{\rm #1}}
\def\b#1{{\bf #1}}
\def\ref#1{$^{#1}$}

\def\m@th{\mathsurround=0pt}
\def\rightarrowfill{$\m@th \mathord- \mkern-6mu \cleaders\hbox{$\mkern-2mu 
\mathord- \mkern-2mu$}\hfil \mkern-6mu \mathord\rightarrow$}
\def\leftarrowfill{$\mathord\leftarrow
\mkern -6mu \m@th \mathord- \mkern-6mu \cleaders\hbox{$\mkern-2mu 
\mathord- \mkern-2mu$}\hfil $}
\def\noarrowfill{$\m@th \mathord- \mkern-6mu \cleaders\hbox{$\mkern-2mu 
\mathord- \mkern-2mu$}\hfil$}
\def\orgate{$\bigcirc \kern-.80em \lor$}
\def\andgate{$\bigcirc \kern-.80em \land$}
\def\inverter{$\bigcirc \kern-.80em \neg$}
\magnification=\magstep1
\tolerance 10000
\baselineskip  12pt
\hoffset=.25in
\hsize 6.00 true in
\vsize 8.75 true in
\font\eightrm=cmr9
\centerline{\bf The Wondrous Design and Non-random Character of ``Chance'' Events}\par\bigskip 
\centerline{Robert A. Herrmann}\parm
\centerline{Mathematics Department}
\centerline{U. S. Naval Academy}
\centerline{572C Holloway Rd.}
\centerline{Annapolis$,$ MD 21402-5002}
\centerline{25 March 1999}\bigskip
{\leftskip=0.5in \rightskip=0.5in \noindent {\eightrm {\it Abstract:} In this article$,$ it is shown specifically that natural system chance events as represented by theory predicted (a priori) probabilistic statements used in such realms as modern particle physics$,$ among others$,$ are only random relative to the restricted language of the theory that predicts such behavior. It is shown that all such ``chance'' natural events are related one to another by a remarkably designed$,$ systematic and wondrous collection of equations that model how the natural laws and processes specifically yield such natural events. A second result shows theoretically that all such ``chance'' behavior is caused by the application of well-defined ultralogics. These results show specifically that the fundamental underlying behavior associated with all natural systems that comprise our universe is controlled internally by processes that cannot be differentiated from those that mirror the behavior of an infinitely powerful mind.
\par}}\par\bigskip

Before I start this article$,$ I must make one very$,$ very important 
disclosure.  All the Herrmann models mentioned and the results they predict utilize  
a few common everyday human experiences that can be verified within a laboratory setting.  Except for one or more of the following principles (1) - (4) and the most basic mathematical axioms used 
throughout all of science$,$ NO additional mathematical or physical axioms or 
presuppositions are assumed. This must always be kept in mind$,$ for there will 
arise the myth that the results have somehow or other been included within the 
axioms or presuppositions. This myth is entirely and utterly false.\parm
From Nobel Laureate Louis 
de Broglie$,$ comes the statement:\parm
{\leftskip=0.5in \rightskip=0.5in \noindent
. . .  the structure of the
material universe has something in common with the laws that govern the 
workings of the human mind. [March$,$ 1963$,$ p. 143]\par}\parm
\noindent In an interesting series of lectures that will be discussed more fully$,$ Nobel Laureate Richard Feynman tells his audience\parm
{\leftskip=0.5in \rightskip=0.5in \noindent
 I am describing {\it how} Nature works. (1985$,$ p. 10)\par}\parm
\noindent These two statements present a rather obvious fact that should come as no surprise. What Feynman and all other materialistic scientists do during their lectures is to describe in words$,$ diagrams$,$ computer images$,$ and the like$,$ the natural laws and processes that they claim will lead to a natural event or a change in a natural system.  For humanly comprehensible ``descriptions,'' as broadly defined$,$ the following is fully discussed in Herrmann (1994) and is denoted as {\bf Principle (1). How nature combines together natural laws and processes to 
produce the moment-to-moment ``evolution'' (i.e. development or changes in the behavior) of a natural system is modeled or mirrored by certain aspects of human mental activity.} One of the most fundamental deductive procedures that models the development of a natural system is$,$ itself$,$  modeled by the ``consequence'' operator. Such operators are discussed in more detail later in this article and many interesting properties appear in Herrmann (1987).  
Principle (1), although clearly restrictive and materialistic in character, is verified, relative to deductive science, by the largest amount of empirical evidence that could ever exist since whenever a scientist predicts natural system behavior from a collection of hypotheses deductive human mental activity is applied. This even includes the application of the principles used to gather and analyze evidence that would tend to verify a scientific hypothesis.    
 \par   
During his 1998 American tour$,$ Stephen Hawking stated while visiting the President at the White House that in a very few years scientists will be able to describe all the natural laws and processes that govern the behavior of every natural system that exists within the universe and use these laws to predict correctly such behavior. This is the basic philosophy of science that drives many scientist. It is immaterial whether or not the Hawking philosophy is correct for the driving force behind {\bf basic research} is that the human brain$,$ let's call it the mind instead$,$ is$,$ at the least$,$ capable of comprehending all the natural laws and processes that govern the workings of our universe. Let's call this  {\bf principle (2)}. \par
I point out that during recent history principle (2) was not accepted by all scientists. For example$,$ Nobel Laureate Max Planck wrote  that:\parm{\leftskip=0.5in \rightskip=0.5in \noindent Nature does not allow herself to be exhaustively expressed in human thought. (1932$,$ p. 2)\par}\parm
Of course$,$   Max Planck is one of the founders of ``Quantum Mechanics.''  Since (2) is not as yet established$,$ it is better to consider (2) as but a partial statement.
Thus$,$ we modify (2) to include the word ``probably$,$'' and accept that some of the present day humanly constructed theories are ``probably'' correct  predictors of objective reality. [Philosophically$,$ the theories themselves should be considered as probable in character (Cohen \& Nagel$,$ 1934$,$ p. 393).] When theories are considered$,$ they will be tacitly assumed to be among the ones that are ``probably'' correct predictors.\par
 In a course on Quantum Information and Computers given at Cal. Tech. by John Preskill we are told in the textbook he wrote for the course that:\parm 
{\leftskip=0.5in \rightskip=0.5in \noindent . . . fundamentally the universe is quantum mechanical. . . . For example$,$ clicks registered in a detector that monitors a radioactive source are described by a {\it truly random} Poisson process. In contrast$,$ there is no place for true randomness in deterministic classical dynamics (although of course a complex (chaotic) classical system can exhibit behavior that is in practice indistinguishable from random). (Preskill$,$ 1997$,$ p. 4)\par}\parm
\noindent {\bf Principle (3) is the notion that fundamental natural system behavior is probabilistic in character.}\par
{\bf [A]} Clearly$,$ Preskill requires his students to take a specific stance in the philosophy of science$,$ a philosophical stance that cannot be scientifically verified in any manner. Although there are deterministic mathematical descriptions for behavior that cannot be distinguished from a pure statistical distribution that it is claim is produced by the ``pure'' random behavior of a natural system$,$ an individual most not accept these deterministic models as reality. I repeat that
according to Preskill and his quantum view$,$ you must not accept these deterministic models as mirroring reality although there is no scientific method that can distinguish such deterministic design for natural system behavior from the claimed non-design displayed by pure random behavior. Of course$,$ we do have the very great mystery why there is$,$ indeed$,$ a Poisson mathematical design being displayed
when a large number of clicks are being considered. But relative to the behavior of each individual click$,$ we are to believe$,$  there is no possible intelligent relationship between the occurrence of one click and the occurrence of the very next click. It is the seeming non-relation between successive clicks that appears to make them 
``random'' in character. \par
First$,$ no one can truly force you to accept this philosophy. You can study chaos theory and other mathematical areas and accept that there are deterministic and highly designed natural world processes being modeled by the mathematics and these natural world processes only give the appearance of random behavior. Of course$,$ I suppose that you would not voice you opinion while a student in such courses at Cal. Tech. But more importantly$,$ does the basic idea of ``random'' imply lack of design? Further$,$ is the apparent random character of quantum mechanics the true underlying feature that describes natural system behavior at all levels of human comprehension?\par
The particle physicists claim that all natural system behavior is fundamentally controlled by the ``invisible'' and individually undetectable entities that abound within the realm of quantum physics and$,$ especially$,$ within an nonempty background that is called the ``vacuum.'' The major evidence that such entities might exist in objective reality is that various theories predict that under certain hypotheses gross matter$,$ that {\bf can} be detected by a machine or a human sensor$,$ will behavior in a certain way. This is called {\bf indirect evidence}. But the entities being described need only be objects within an {\bf analog model}$,$ a collection of fictitious entities that are used solely to aid  the human mind to comprehend and describe processes that will predict an experimental outcome although the entities and processes need not exist in objective reality.\par
Many times the subatomic entities start out as imaginary. In the original paper Einstein wrote where he gives his model for the photoelectric effect$,$ he called the photon an ``imaginary'' particle-like entity. But$,$ Richard Feynman in his book ``QED: The Strange Theory of Light and Matter'' (1985) insists that light particles exist in objective reality. However$,$ in this same book in order to explain partial reflection Feynman states relative to the direction of an arrow that will measure a probability:\parm
{\leftskip=0.5in \rightskip=0.5in \noindent To determine the direction of each arrow$,$ let's imagine that we have a stopwatch that can time a photon as 
it moves. This imaginary stopwatch has a single hand that turns around very$,$ very rapidly. When a photon leaves a source$,$ we start the stopwatch. As long as the photon moves$,$ the stopwatch hand turns . . . ; when the photon ends up at the photomultiplier$,$ we stop the watch. The hand ends up pointing in a certain direction. This is the direction we will draw the arrow. (Feynman$,$ 1985$,$ p. 27.)\par}\parm
I point out that not only is the stopwatch imaginary$,$ at least at the present time$,$ but the vectors$,$ the arrows$,$ also do not exist in objective reality. Such descriptions form but an analog model for such behavior and nothing more than that. Indeed$,$ many physicists in this area of QED use such phrases as ``a photon is absorbed'' or ``emitted by an electron'' without ever considering any description as to how an ``electron'' absorbs or emits anything. An assumed process such as the directly unobserved interaction between a photon and an electron is an essential part of this QED theory. But$,$ not withstanding these obvious logical faults$,$ I will consider as another principle of modern science$,$ {\bf principle (4)$,$ the principle of indirect verification} as fundamental to modern research. This principle states that if assuming the existence of undetectable entities or processes yields correct predictions for the behavior of gross matter$,$ then these entities or processes should be assumed to exist in objective reality. Of course$,$ principle (4) immediately implies that principle (1) is empirical fact. Also note that from a theoretical point of view$,$ the fewer hypotheses a scientist employs as a bases for a theory$,$ the better the theory is said to be.\par
 Assuming principle (2) and (4)$,$ human beings describe laws and processes of nature as part of human theories and predict correctly from these theories natural system behavior$,$ even if all such predictions are probabilistic in character.  Consequently$,$ the following must hold. Let $\Gamma$ be a set of  hypotheses of such a theory. The $\Gamma$ need not contain all of the natural laws or processes that will lead to a prediction$,$ but need only contain a description for the physical entities to which these natural laws or processes will be applied. Then the human brain takes $\Gamma$ and preforms upon $\Gamma$ 
some sort of mental process that yields a prediction $P$. This prediction is then verified in the laboratory or$,$ similar to the Big Bang cosmology$,$ is assumed to have occured sometime in the past.  These$,$ at present$,$ unknown human mental processes are indicated by the {\bf turnstile} symbol $\vdash.$
If $\Gamma$ does not contain all of the applicable natural laws or processes$,$ then these laws or processes become part of the human mental process symbolized by $\vdash.$  
In either case$,$ this simple sequence of events is symbolized by  
$$\Gamma \vdash P.\eqno [1]$$\par
In all that comes next$,$ the fundamental hypotheses will include the previously mentioned four principles. These principles are re-stated as follows:\parm
{\leftskip=0.5in \rightskip=0.5in \noindent 
(1) How nature combines together natural laws and processes to 
produce the moment-to-moment ``evolutio'' (i.e. development or changes in the behavior) of a natural system is modeled or mirrored by certain aspects of human mental activity.\par
\noindent (2) Some of the present day humanly constructed theories are ``probably'' correct  predictors of objective reality. When theories are considered$,$ they will be tacitly assumed to be among the ones that are ``probably'' correct predictors.\par
\noindent (3) Predictions of correct fundamental natural system behavior will always require the behavior to be associated with a statistical statement that implies that natural system behavior originates from behavior that is always probabilistic in character.\par
\noindent (4) Although natural entities and processes described by means of a language using ``physical'' terms may not be observed directly$,$ they  will be considered as existing in objective reality if their continued use yields correct predictions.\par}\parm  
It is self-evident that according to the principle (2) that [1] displays an absolute fundamental process  that occurs within the human brain. Expression [1] corresponds to principle (1) in that it displays a fundamental relationship between human mental processes and natural system behavior. Expression [1] and the next expression [2] can also be related to the {\bf informational content} of a theory and Preskill claims that information is ``physical'' (Preskill$,$ 1997$,$ p. 4) in character.  Mathematicians study patterns associated with human mental activity in various ways. There are general properties associated with these mental patterns and these mental properties can be modeled by means of a mathematical object called  a {\bf consequence} operator $C$
rather than the turnstile $\vdash.$ Relative to a consequence operator the turnstile relation becomes 
$$P \in C(\Gamma).\eqno [2]$$

Expression [2] displays$,$ in the simplest possible manner$,$ the possible existence of a natural process that is being mirrored by the properties of operator $C$.  Since the consequence operator properties are fundamental to all physical science theory and principle (2) states that some of the probably correct humanly generated theories are known and these theories predict the behavior of natural systems within the universe$,$ then using principle (4) we should accept that it is absolutely true in objective reality that the behavior of operator $C$ is mirroring the most fundamental of all natural processes. \par
One of the major characteristics of a modern scientific community is that it is exceptional myopic when it deals with scientific theory. A scientist learns and then concentrates upon a very complex often mathematically based theory and predicts verified results. But$,$ usually$,$ the scientist cannot step back and observe the actual essence of what has been accomplished through such efforts. What is actually happening in nature itself is almost always hidden among thousands of pages of symbolic representations. This also happens when mathematicians study the patterns presented by consequence operators as a model for what$,$ using principle (1)$,$ must be natural processes. \par  

 A consequence operator is defined upon all of the subsets of a language $\cal L$ and $\cal L$  can be very broad in character and not only include the ordinary strings of symbols one associates with a language such as those strings displayed in this article but can include audio and visual impressions as well. You need not start with a turnstile operator $\vdash$ as your basic operator. It has been customary to include in the $\Gamma$ statements for all of the theoretical aspects that the human mind uses to predict an output event that should occur when a describable natural system
input is being considered. For example$,$ the natural input for radioactive decay is the presence of a ``decaying'' radioactive material.  It should be possible to apply certain general natural laws and processes that explain such decay and$,$ of course$,$ application of human mental processes$,$ and predict the probability$,$ with respect to time intervals$,$ of the occurrence of the detector ``clicks.'' Then as another example$,$ using the Feynman described theory for partial reflection$,$ the actual source is a ``photon generator'' and the detectors are photomultipliers. As Feynman states it on page 17 of the above mentioned book$,$ QED will predict that for a glass surface and a supply of photons of a fixed frequency$,$ say red light$,$ on the average 4\% of the photons out of 100 photons will be reflected by the surface at a specific angle and be counted by a specially placed photomultiplier. The input or {\bf trials} are the 100 photons emitted from the source and the output events are the 4 photons detected by the photomultiplier. However$,$ as mentioned$,$ one need only consider $\Gamma,$ for the radioactive case$,$ as a single description for the radioactive material$,$ and$,$ for the photon case$,$ a description for the photon source.  \par 
With respect to these event predictions$,$ the essence of the concept of random is that the occurrence of one of these events has no influence upon the occurences of the ``next event.'' (Olkin$,$ et al. 1994$,$ p. 7.) The basic problem is what does the phrase ``no influence'' mean. One of the aspects of the new material to be presented is that the phrase ``no influence'' does not mean no design.  Further$,$ it will be shown that the phrase ``no influence'' relative to predictions produce by any physical theory means {\bf [B] no influence that can be described using the theory's restricted language (Theory Randomness)}$,$ with the exception of trivial identity styled relations. But$,$ although principle (2) is used by most scientists to mean ``no influence'' by means of any natural law or process$,$ which is the concept called {\bf absolute randomness}$,$ we will show that there is no such scientific concept as  absolute randomness. Indeed$,$ absolute randomness  is a false statement.  We will show that the myopia of the scientific community has prevented it from ``seeing,'' so to speak$,$ the actual wondrous design and the actual basic event influences that most be present before any so-called random events can occur. \par  
In explaining the following new research findings$,$ certain philosophical statements made by Feynman will be upheld. Relative to theoretic constructs$,$ Feynman writes:\parm
{\leftskip=0.5in \rightskip=0.5in \noindent The . . . reason that you might not understand what I am telling you is$,$ while I am describing {\it how} Nature works$,$ you won't understand {\it why} Nature works that way. But you see$,$ nobody understand that. (1985$,$ p. 10)\par}\parm  
\noindent From my experience$,$ another aspect of the Feynman philosophy is perfectly correct with the exception of$,$ at the least$,$ one of notable research discovery.\parm
{\leftskip=0.5in \rightskip=0.5in \noindent Finally$,$ there is the possibility: after I tell you something$,$ you just can't believe it. You can't accept it. A little screen comes down and you don't listen anymore. I'm going to describe to you how Nature is -- and if you don't like it$,$ that's going to get in the way of your understanding it. It's a problem that physicists have learned to deal with: They've learned to realize whether they like a theory or they don't like a theory is {\it not} the essential 
question. Rather$,$ it is whether or not the theory gives predictions that agree with experiment. It is not a question of whether a theory is philosophical delightful$,$ or easy to understand$,$ or perfectly reasonable . . . . (1985$,$ p. 10)\par}\parm
The notable exception is a {\bf theory of everything} (Herrmann$,$ 1994) that solves the General Grand Unification problem. This theory uses processes that satisfy$,$ in general$,$ principles (1)$,$ (4)$,$ and since it predicts the behavior of all of the natural systems that exist within our universe$,$ it should$,$ by principle (4)$,$ be accepted by all of the scientific community. In general$,$ the model shows$,$ with respect to principle (4)$,$ that our universe was created by processes that mirror the processes one would associate with an infinitely powerful mind. The model is called the MA-model and uses the processes and concepts associated with a mathematical entity called the Nonstandard Physical World which also$,$ according to principle (4)$,$ should be accepted as existing in objective reality.  The fact that most of the  scientific community rejects this model is a counterexample to the last sentence in the above Feynman quotation. Since the MA-model does depend upon the existence of a ``background'' or ``substratum'' universe that is the domain for the universe creating operators$,$ then$,$ counter to principle (4) and the Feynman statement$,$ this model is further rejected based upon the philosophical stance that no such undetectable substratum exists.
However$,$ postulating away the existence of such a background universe is not sufficient in order to eliminate the conclusion that natural systems within our universe are$,$ indeed$,$ controlled by processes that mirror exceptionally
remarkable and wondrous mental processes. \par 
As final re-enforcements to the above four principles$,$ where I have added remarks between the [ and ]$,$ Feynman states:\parm
{\leftskip=0.5in \rightskip=0.5in \noindent I am not going to explain how the photons actually ``decide'' when to bounce back or go through: this is not known . . . . I will only show you how to calculate the correct [Indeed$,$ as perfect as one wishes]  
{\it probability} that light will be reflected from the glass of a given thickness$,$ because that's the only thing physicists know how to do! What we do to get the correct answer to {\it this} problem is analogous to the things we have to do to get the answers to {\it every other} problem explained by quantum electrodynamics [QED]. (1985$,$ p. 24)\parm
\noindent The theory [QED] describes {\it all} of the phenomena of the physical world except the gravitational effects . . . and radioactive phenomena [nuclear physics] . . . [ Note: recent research tends to uphold the original Hawking principle in that procedures similar to those used for QED seem to  predict some gravitational and many nuclear effects.]  Most phenomena we are familiar with involve such {\it tremendous} numbers of electrons that it's hard for our poor minds to follow that complexity. [But$,$ in theory$,$ a computer$,$ which follows the specific logical rules of the propositional logic$,$ should be able to make such calculations.] (1985, p. 7-8)\par}\parm 
The following discussion and results$,$ although applicable to various physical scenarios$,$ are being restricted to the realm of ``particle physics.''
All probability statements can be re-expressed in terms of the number $n$ of describable {\bf trials} and specific describable natural events \b A that can occur during these specific trials. Since all of these theories use mathematical procedures to predict a probability $p,$ the basic requirement for a probability statement is that the trials form a 
{\it random sequence} under a given set of conditions. If during the $n$ trials$,$ where $n$ is a large number of trials$,$ $m$ events \b A occur$,$ then the event \b A occurs with a probability $p$ approximated by $m/n.$ The approximation becomes more certain as the number of trials increases ``without limit.'' With the exception that one might claim$,$ as Preskill has done$,$ that certain events seem to follow a mathematical distribution$,$ the ``random sequence'' statement cannot be established within a laboratory setting with certainty. As pointed out in the dictionary by James and James (1968, p. 285 p. 307)$,$ {\it such definitions have either logical or empirical difficulties.}\par
Of course$,$ any description for the source that produces the trials$,$ the trials themselves and a description for the events requires a specific scientific language. Note that some trial descriptions are solely relative to ``time.'' One calculates that over certain time intervals the probability that the detector will click is such and such. On the other hand$,$ you might have a collection of different events that can be described in the appropriate language. For example$,$ suppose that out of 100 photons the probability that they will be ``reflected'' at the measured angle is 0.40 (i.e. event \b A occurs) and the probability that they will be ``scattered'' or not reflected at the measured angle is 0.96 (i.e. event \b A does not occur). Each trial $n$ and each event produced by a trial must correspond to some sort of ``counting label'' or else you would have no such probability statement. \par
Since mental activity is used to predict probability notions$,$ the $n$ trials can be modeled by means of $n$ very simple consequence operators$,$ one for each of the counting labels and the $n$ possible outcomes$,$ of which $m$ are the \b A events. Let $\Gamma$ be a description for the source such as ``A photon from S.'' Let each $b \in B,$ where $\vert B\vert\geq 2,$ correspond to a description for the event outcome produced by each of the $n$ trials. There needs to be $n$ applications of a mentally produced statement such as ``$\Gamma$ yields a member of $B$.'' In the case of  \b A events$,$ $m$ of these trials correspond to a fixed $b \in B$ that describes the \b A events. Using a power set map$,$ the following set of statements model the essence of this probabilistic scenario$,$ where for each $i$ such that $1\leq i \leq n$ there is some $b\in B$ such that $b = a_i.$ 
$$a_1 \in H_1(\{\Gamma\}), \cdots , a_n\in H_n(\{\Gamma\}),\eqno [3]$$
where the counting label is represented by the $n$ applications of the operator $H$ (as denoted by the $H_1,\ldots, H_n$ symbols) as $H$ is associated with the labeled $a_1,\ldots, a_n$ events. \par
In order to indicate the intuitive ordering of any sequence of events, the set $\r T$ of Kleene styled ``tick'' marks is used (Kleene$,$ 1967$,$ p. 202) as they might be metamathematically abbreviated by symbols for the non-zero natural numbers.
Using the philosophy of science concept of simplicity$,$ assume that our language ${\rm L} = \{\Gamma\} \cup \r B \cup \r T.$ In 1987$,$ it was discovered that there exists a set of consequence operators $${\cal C}= \{C_i=C(\{a_i\},\{\Gamma\})\mid 1 \leq i \leq n \},\eqno [4]$$ 
that satisfies the requirements of [3] and is minimal with respect to $\rm L$ and [3] (the Occam Razor requirement). (Herrmann$,$ 1987$,$ p. 2$,$ Definition 2.4 (i). Note: This 1987 paper$,$ unfortunately$,$ contains numerous topographical printers errors.) Further$,$ in order to maintain the requirement and$,$ hence$,$ prediction that {\bf [C] no two or more of the trial   
events $b$ be in a recognizable order (an aspect of theory randomness)}$,$ principle (1) requires that any two or more members of ${\cal C}$ in expression [4]$,$ when combined together$,$ be a simple type of consequence operator such that the corresponding event outputs of this combination maintain this requirement. 
The appropriate combination that is absolutely necessary is called the {\bf union} consequence operator and is denoted by $C'= C_1 \vee C_2 \vee \cdots \vee C_n$ (i.e. $C'(X) = C_1(X) \cup \cdots \cup C_n(X)$). The union consequence operator does not usually exist$,$ but in the case of the set of operators in ${\cal C}$ such an operator defined on $\rm L$ does exist and is minimal in the sense of it being the weakest consequence operator (mental activity) that is stranger than each $C_i$ and$,$ hence$,$ satisfies the Occam Razor requirement for models (Herrmann$,$ 1987$,$ p. 5). Such consequence operators are the most simplistic entities one can use to model the actual mental processes that yield the probabilistic predictions.\par
 There is another minimal mental-like operator that would be applied after each $C_i$. In human communication$,$ one of the most significant processes is the selection from a vast collection of words and phrases a particular collection that faithfully describes an event. This operator is called the {\bf finite human choice} operator. This operator mirrors the natural process known as the {\bf realism} operator and might be considered as a slight generalization of the natural selection operator used throughout  evolutionary theories. However$,$ this operator is usually considered as but an integral part of each $C_i.$ This operator$,$ denoted by $R$$,$ is restricted to members of $\cal C$ as they are applied to $\{\Gamma\}$ and such operators as $C'.$ Define $R(C(\{\Gamma\}) = C(\{\Gamma\}) - \{\Gamma\}.$ This yields for each (trial) $R(C_i(\{\Gamma\})) = \{a_i\}$ and $R(C'(\{\Gamma\}) = \r B.$\par
A source $\Gamma_1$ can have numerous associated but different $\r B_j,\ 1 \leq j \leq k$$,$ \b A described events probabilistically predicted by a specific theory. For example$,$ simply consider the source as sun-light. Again  these probabilistic predictions correspond to a set of consequence operators ${\cal C}_1$ each defined on a language ${\rm L}_1 = \{\Gamma_1\} \cup \r B_1 \cup \r B_2 \cup \cdots \cup \r B_k \cup \r T.$ Again these consequence operators must satisfy the union operator requirement. The simplest set of such consequence operators that contains ${\cal C}_1$ and satisfies the union operator property forms one of the most wondrous and beautifully designed of all mathematical objects: it is a complete distributive lattice of finitary consequence operators (Herrmann$,$ 1987$,$ p. 5). [For reference purposes$,$ the set is ${\cal C}_2= \{C(X,\{\Gamma_1\})\mid X \subset {\rm L}_1\},$ where if $\Gamma_1 \in Y,$ then $C(X,\{\Gamma_1\})(Y) = Y \cup X;$ if $\Gamma_1 \notin Y,$ then $C(X,\{\Gamma_1\})(Y) = Y.$] \par
Usually$,$ the probability statement is considered to be more accurate only if you increase the number of trials $n$ without limit (i. e. $n \to \infty$). In order to avoid the philosophical difficulties one might have with the ``infinity'' concept$,$ one can use the ``potential'' infinite notion. In this case$,$ one can describe this process by saying$,$ ``Pick any natural number you wish$,$ then the necessary number of trials for an accurate prediction is greater than your choice.'' The conclusion that the set of all such consequence operators has this remarkable mathematical property is not dependent upon the number of trials.\par
Such sets of consequence operators have a much more startling property$,$ however. Any set of consequence operators such as those in [4] and that satisfies the requirement that $C'$ is a consequence operator are related one to another. In the appendix is a new nontrivial result that shows that for any nonempty set of consequence operators $\{C_i \mid 1\leq i \leq n,\ 1 < n\},$ each of which is defined on a language $\cal L$ and for which $C'$ is a consequence operator defined on $\cal L$$,$ it follows that 
$$C_1(C_2 \vee \cdots \vee C_n) = C_2(C_1 \vee C_3 \vee \cdots \vee C_n)= \cdots$$
$$C_{n-1}(C_1 \vee \cdots\vee C_{n-2}\vee C_n)= C_n(C_1\vee C_2 \vee \cdots \vee C_{n-1}). \eqno [5]$$\par  
Although$,$ there may not be a relation between individual events $a_i$ that can be described in terms of the restricted language used for the predictive physical theory$,$ there is the relation [5] between each of the individual operators described above for a probabilistic model and that are needed to predict each $a_i.$ Using principle (1)$,$ the consequence operators $C_i$ correspond to natural processes that combine together the theory dependent natural laws or processes in order to predict the probabilistic statement. Hence$,$ the natural processes modeled by the $C_i$ are individually related by the $n(n - 1)/2$ equations that appear in [5]. Expression [5] is what one might describe as being in a symmetric form. But$,$ actually$,$ there are many more such equations since the operator $\vee$ is commutative and$,$ hence$,$ any permutation of the operators between the $($ and $)$ will also lead to many more such equations. Consequently$,$ the events $a_i$ that are produced by the necessary consequence operators $C_i$ are not absolutely random in character when the processes are described by the theory of consequence operators. This also indicates that the actual concept of ``randomness'' should be considered as only relative to the original theory restricted language that predicts the probabilistic statement.\par
It is self-evident from paragraph {\bf [A]} on page 2$,$ the theory randomness statement {\bf [B]} on page 6$,$ and statement {\bf [C]} relative to recognizable order on page 9 that such ``randomness'' need not be considered as a hypothesis within such probabilistic theories as QED. Indeed$,$ theory randomness is a statement about and$,$ hence$,$ exterior to such theories and as such is a statement within the employed philosophy of science. For this reason$,$ expression [5] is falsifiable. For if it can be demonstrated that {\bf [C]} is false$,$ then this will falsify the expression [5]. \parm 
\centerline{\bf Conclusions}\parm
As mentioned$,$ the MA-model shows that it is rational to assume that an external mental-like consequence operator and an external object that behaves like an actual collection of symbols$,$ a word$,$ are the underlying entities that produce and sustain not only our universe in its evolutionary development but can be used to create and sustain other ``universes'' as well. 
The particle physics community tends to accept the original Hawking principle$,$ especially since their contention is that it is the probabilistic rules and processes of quantum physics that will be shown$,$ shortly$,$ to govern completely the four fundamental interactions produced by what are often called the electromagnetic$,$ strong$,$ weak$,$ and gravitational ``forces.''
Due to the often stated requirement that probabilities be considered as related to potentially infinitely many trials$,$ many would consider the $n$ that appears in displays [4]$,$ [5] to be potentially infinite in character. Notwithstanding this last possible requirement$,$ since the $n$ mental-like consequence operators that appear in [4] and [5] must apply to every predictive quantum physical scenario in the vicinity of every spacetime location throughout our universe$,$ as assumed by particle physics$,$ then the fundamental behavior that governs our universe can be described$,$ in general$,$ in the exact same terms as those used to describe the MA-model conclusions. Consequently$,$ it appears that the fundamental underlying behavior associated with all natural systems that comprise our universe is controlled internally by processes that mirror the behavior of an infinitely powerful mind.
 \par
The research result Theorem 2$,$ in the appendix$,$ gives a very significant additional result relative to the infinitely powerful mind concept.
Relative to quantum physical behavior of photons and partial reflection as QED predicates the probability that an event will occur$,$ Feynman writes:\par 
{\leftskip=0.5in \rightskip=0.5in \noindent I am not going to explain how the photons actually "decide" whether to bounce back or go through; that is not known. (Probably the question has no meaning.) (1985$,$ p. 24)\par}\par
Feynman uses at this point in his lecture the terms ``bounce back'' and ``go through'' for what he later terms as new photons (emitted from electrons) and that reach the detector or photons that are ``scattered'' by electrons. The probability that an observed event will occur depends upon the {\bf combined} probability that a set of individual events will occur. Since natural systems are also controlled$,$ in general$,$ by ultrawords$,$ it would be the ultranatural events and ultrawords that force the individual probabilities to combine in the appropriate manner that would be described as quantum physical ``natural law.'' Consequently$,$ this combined probability depends upon the probability that a describable individual event \b E will occur.\par
 The above Feynman statement may$,$ indeed$,$ be correct for the language of QED and quantum mechanics in general$,$ but it is a false statement relative to the theory of ultralogics and the physical-like behavior of ultralogics. Relative to theory predicted probabilistic behavior, Theorem 2$,$ in the appendix$,$ shows explicitly that for any sequence of relative frequencies $m/n$ that converges to a predicted probability $p$ that an individual event will occur$,$ whether the convergence be fast or slow$,$ there is a physical-like ultralogic that does$,$ indeed$,$ force the event to occur or not to occur in such a manner that the relative frequence sequence is duplicated exactly. Accept for satisfying the basic consequence  operator properties for what are called internal sets$,$ it is significantly$,$ that the ultralogic $P_p$ that forces a specific natural system to so comply with such a probabilistic pattern is {\bf not} related to any known human deductive process. \par
Preskill's (1997$,$ p. 4) quoted remark is about a probability distribution, the Poisson distribution. The facts are that all such distributions that are obtained from a frequency function follow  patterns dictated by an ultralogic generated by a finite set of ultralogics $P_{p(i)}$. Relative to Theorem 2$,$ the only difference  
is that the ``source'' statement $\Gamma$ maybe more complex in that it describes a particular physical scenario and the ``event'' statemens in $\rm B$ may also be more complex. Under the view that equivalent descriptions represented by $\Gamma$ and members of $\rm B$ contain all the necessary information that depicts natural-system behavior for a particular physical scenario, then a frequency function $f(x)$ is used to obtain the probability that the scenario will occur. A product consequence operator (an ultralogic) is generated by finitely many of the $P_p$ that appear in Theorem 2 and yields the appropriate sequence of events that satisfies the required distribution. 
\par
Using these event sequence ultralogics$,$ one would conclude that the so-called ``unregulated or random'' behavior that one often associates with quantum mechanics and the behavior of entities within our universe is actually one of the most powerful signatures that processes represented by an infinitely powerful mind description are controlling all aspects of natural system behavior.\parm \vfil\eject   
\centerline{\bf Appendix}\parm
Prior to presenting the major mathematical results$,$ one technical aspect for such modeling needs to be explored. This aspect was assumed to be self-evident in some of the previous Herrmann articles written on this subject. However$,$ this ``self-evidence'' will now be further detailed. Different informal (i.e. native) languages use different discipline dictionaries as well as different syntactic rules. In formal logic$,$ the same symbols and same syntactic rules yield different strings of symbols that are said to be ``logically equivalent.'' Throughout various deductive arguments logically equivalent strings of symbols can be substituted$,$ one for another$,$ and the logical argument remains valid. This type of equivalence is defined in a strict and not controversial manner. Types of equivalence are necessary when informal language descriptions are used. However$,$ a definition as to the equivalence of different native language strings of symbols is necessarily more intuitive in character. \par
First note that to comprehend properly a set of symbol strings that contains more than one string of symbols it is usually required that the strings to be presented in a specific order. When necessary$,$ the use of a spacing and punctuation symbols would allow any finite set of such symbol strings to be considered as a set that contains but one string of symbols. However$,$ even under  a single symbol string  notion$,$ various types of equivalence are required for informal language descriptions. Intuitively$,$ this equivalence is captured within the statement that ``two strings of symbols are {\bf saying the same thing} or have the {\bf same relative meaning.}'' It is necessary that a form of mental activity and general consensus be considered in order to come to this conclusion. Whatever technique is used   
for this purpose$,$ such a concept appears to have universal acceptance or else it would not be possible to ``translate'' efficiently from one native language to another and ever hope to achieve the same intended results being communicated by two different strings of symbols. Of course$,$ the same idea of equivalent symbol strings would hold within a specific meaningful language as well. \par
 The set of all ``equivalent'' strings of symbols taken from the language and that have been adjudged to be ``saying the same thing'' is an {\it equivalence class.} Distinct equivalence classes are disjoint$,$ that is they have no members in common$,$ and every member of the language is contained in some equivalence class. The self-evident notion previous made but not explicitly stated is that the actual set that determines the domain and codomain for our consequence operators is a set composed of one and only one member$,$ a {\it representative}$,$ from each of these equivalence classes. When dealing with descriptions using a native language$,$ this procedure is tacitly assumed.\par 

Let $\cal L$ be a nonempty language and ${\cal P}$ denote the power set operator. A mapping $C\colon \power {\cal L} \to \power {\cal L}$ is a general consequence operator if for each $X \in \power {\cal L}$ (1) $X \subset C(X),$ (2) $C(C(X))=C(X).$ (3) If $X,Y \in \power {\cal L}$ and $X \subset Y,$ then $C(X) \subset C(Y).$ \parm\vfil\eject
THEOREM 1. {\it Let $\cal C$ be the set of all consequence operators defined on ${\cal L}.$ Let $C_1,\cdots C_n\in {\cal C},\ n >1,$ and$,$ for each $X \subset {\cal L},$ let $(C_1 \vee C_2\vee \cdots \vee C_n)(X)= (\bigvee_{i=1}^n C_i)(X)= \bigcup_{i=1}^n C_i(X)$. If $\bigvee_{i=1}^n C_i\in \cal C$$,$ then 
$$ C_i(\bigvee_{j\in\{1,\ldots,n\}-\{i\}}C_j)=C_k(\bigvee_{j\in\{1,\ldots,n\}-\{k\}} C_j),\ 1\leq i,k\leq n$$.}\par\noindent
Proof. Define for each $X \subset {\cal L}, \ (\bigvee_{i=1}^n C_i)(X) = C_1(X) \cup \cdots\cup C_n(X).$
Notice that $\bigvee_{i=1}^n C_i$ is a mapping on the power set of ${\cal L}$ into the power set of ${\cal L}$ and satisfies axioms 1 and 3 for consequence operators. Two auxiliary results need to be established using our three axioms. \par
Suppose that $X,\ Y \subset {\cal L}$ and $C \in \cal C.$  
From axiom 1$,$ we have that $X \cup Y \subset X \cup C(Y) \subset C(X) \cup C(Y).$ Application of axiom 3$,$ yields that [1] $C(X \cup Y) \subset C(X \cup C(Y)) \subset C(C(X) \cup C(Y)).$ Since $X \subset X \cup Y$ and $Y \subset X \cup Y$$,$ axiom 3 yields that $C(X) \subset C(X \cup Y),\ C(Y) \subset C(X \cup Y).$  Consequently$,$ $C(X) \cup C(Y) \subset C(X \cup Y).$ Applying axiom 2$,$ it follows that [2] $C(C(X)\cup C(Y)) \subset C(C(X\cup Y)) = C(X \cup Y).$ It follows from [1] and [2] that [3] $C(X \cup Y) = C(X \cup C(Y)) = C(C(X) \cup C(Y)).$\par
For a given $C \in \cal C$$,$ a $Y \subset {\cal L}$ is called a {\it C-system} if and only if $C(Y) \subset Y, $ which is equivalent to $C(Y)=Y.$ 
Note that for any consequence operator $C\in \cal C$$,$ ${\cal L}$ is a C-system. Let $ S(C)$ be the nonempty set of all $C$-systems for a given $C\in \cal C.$ Suppose that  [4] $\emptyset \not= {\cal A} \subset S(C)$ and [5] $X= \bigcap\{Y\mid Y \in {\cal A}\}$. Let arbitrary $Y \in {\cal A}.$ Then $C(Y) \subset Y.$ But fixed $X \subset Y$ implies that $C(X) \subset C(Y),$ which implies that $C(X) \subset \bigcap \{C(Y)\mid Y \in {\cal A}\}= \bigcap \{Y\mid Y \in {\cal A}\}= X.$  Hence [6] $X\in S(C).$\par
For $C_1,\ C_2 \in \cal C$ define $C_1 \leq C_2$ if and only if
for each $X \in {\cal L},\ C_1(X) \subset C_2(X).$ The binary relation $\leq$ is a partial order on $\cal L.$ The algebra $\langle {\cal C}, \leq \rangle,$ along with other objects and relations is shown by W\'ojcicki (1970) to be a complete lattice. Our interest is in the structure of the least upper bound (the supremum) $C_0$ for the $C_i \in {\cal C},\ i = 1,\ldots,n.$ As shown by W\'ojcicki (1970$,$ p. 276) for any $X \subset {\cal L},\ C_0(X) = Y_X = \bigcap \{Y\subset {\cal L} \mid X\subset Y=C_1(Y)=C_2(Y)=\cdots C_n(Y) \}.$ Thus the $Y_X$ is by [6] a $C_i$-system $i = 1, \ldots n.$ Indeed$,$ intuitively
the smallest (with respect to $\subset$) such common $C$-system. Now
$X \subset Y_X$ implies that $X \subset C_i(X) \subset C_i(Y_X)= Y_X,\ i=1,\ldots,n.$ Consequently$,$ $(\bigvee_{i=1}^n C_i)(X)\subset Y_X.$ Further$,$ $C_i(X) \subset C_1(X) \cup \cdots \cup C_n(X),\ i=1,\ldots,n.$ Assume that $\bigvee_{i=1}^n C_i\in \cal C.$ First$,$ notice that 
$C_i\leq \bigvee_{i=1}^n C_i, \ i=1,\ldots,n,$ and $\bigvee_{i=1}^n C_i\leq C_0.$ Thus $\bigvee_{i=1}^n C_i= C_0$ and satisfies axiom 2. This implies that for arbitrary $X\subset {\cal L}, \ (\bigvee_{i=1}^n C_i)(X)=Y_X=C_1(Y_X)= C_2(Y_X)=\cdots =C_n(Y_X)$ by [6]. Therefore$,$ the composition  $(C_i(\bigvee_{i=1}^n C_i))(X)=C_i(Y_X)=C_k(Y_X)=
(C_k(\bigvee_{i=1}^n C_i))(X),\ 1\leq i,k\leq n.$ 
From [3]$,$ for each $i$ such that $1\leq i\leq n,$ $(C_i (\bigvee_{i=1}^n C_i))(X)= C_i(\bigcup_{i=1}^n C_i(X))=
C_i(X \cup (\bigcup_{j\in \{1,\ldots,n\}-\{i\}} C_j(X)))=C_i(\bigcup_{j\in \{1,\ldots,n\}-\{i\}} C_j(X))= (C_i(\bigvee_{j\in \{1,\ldots,n\}-\{i\}} C_j))(X),$ since $X\subset \bigcup_{j\in \{1,\ldots,n\}-\{i\}} C_j(X),$ and the result follows.\parm
[Theorem 1 also holds for any set of finitary consequence operators defined on $\cal L$. by Herrmann (2003).]\parm
In what follows$,$ although this is not necessary$,$ we will be more specific and restrict our attention to the sets and objects discussed within the main body of this paper. Moreover$,$ any new terminology is taken from Herrmann (1993).\parm
THEOREM 2. {\it For the language {\rm L} and any $p\in \real$ such that $0\leq p\leq 1,$ where $p$ represents a theory predicted (i.e. a priori) probability that an event will occur$.$ There exists choice function $C$  and an ultralogic $ P_{p}$ with the following properties.\par
$1.$ When $P_{p}$ is applied to $\{\b G \},\ \r G = \Gamma,$ a hyperfinite set of events $\{a_1,\ldots,a_n,\ldots,\kern-.3em\hyper a_{\nu} \}$ is obtained such that for any ``n'' trials$,$  $\{a_1,\cdots,a_n\}$ is a finite identified event sequence$,$ where each $a_i$ satisfies the labeled event statement $\b E$ or labeled non-event statement $\b E'$ predictions.\par 
$2.$ The predictions in $1$ are sequentially determine by $C,$ where $C$ determines a sequence $g_{ap}$ of relative frequencies that converges to $p$ as the events are directly or indirectly experimentally verified$.$ \par
$3.$ The sequence of relative frequencies $g_{ap}$ determined by $C$ gives the appearance of theory dependent random chance.}\par\bigskip\noindent
Proof.  
 All of the objects discussed will be members of an informal superstructure at a rather low level and slightly abbreviated definitions$,$ as also discussed in Herrmann (1993$,$ p. 23$,$ 30-31)$,$ are utilized. As usual $\nat$ is the set of all natural numbers including zero$,$ and $\nat^{> 0}$ the set of all non-zero natural numbers.\par
Let $A=\{a\mid (a\colon \nat^{>0} \to \nat)\land(\forall n(n \in \nat^{>0} \to (a(1)\leq 1\ \land\  0\leq a(n+1)-a(n)\leq 1)))\}.$ Note that the special sequences in $A$ are non-decreasing and for each $n \in \nat^{>0},\ a(n)\leq n.$ Obviously $A \not=\emptyset,$ for   
the basic example to be used below$,$ consider the sequence $a(1) =0, \ a(2) =1,\ a(3) =1, \ a(4) =2,\ a(5) =2,\ a(6) = 3,\ a(7)= 3, \ a(8) = 4, \ldots$ which is a member of $A.$ Next consider the must basic representation $Q$ for the non-negative rational numbers
where we do not consider them as equivalence classes.   Thus 
$Q = \{(n,m)\mid (m \in \nat)\land (n \in \nat^{>0})\}.$  \par
For each member of $A,$ consider the sequence $g_a \colon \nat \to Q$ defined by $g_a(n) = (n,a(n)).$ Let $F$ be the set of all such $g_a$ as $a \in A.$ Consider from the above hypotheses$,$ any 
$p \in \real$ such that $0\leq p \leq 1.$ We show that for any such $p$ there exists an $a \in A$ and a $g_{ap} \in F$ such that $\lim_{n \to \infty}g_{ap}(n) = p.$ For each $n\in \nat^{> 0},$ consider $n$ subdivision of $[0,1],$ and the corresponding intervals $[c_k,c_{k+1}),$ where $c_{k+1} - c_k =1/n,\ 0\leq k < n,$ and $ c_0=0,\ c_n = 1.$ If $p=0,$ let $a(n) = 0$ for each $n \in \nat^{> 0}.$  Otherwise$,$ using the customary covering argument relative to such intervals$,$ the number $p$ is a member of one and only one of these intervals$,$ for each $n \in \nat^{> 0}.$ Hence for each such $n>0,$ select the end point $c_k$ of the unique interval $[c_k,c_{k+1})$ that contains $p.$ Notice that for $n = 1,$ $c_k = c_0 = 0.$ For each such selection$,$ let $a(n) = k.$ Using this inductive styled definition for the sequence $a,$ it is immediate$,$ from a simple induction proof$,$ that $a \in A,\  g_{ap} \in F,$ and that $\lim_{n \to \infty}g_{ap}(n) = p.$ For example$,$ consider the basic example $a$ above. Then $g_{ap} = \{(1,0),(2,1),(3,1),(4,2),(5,2),(6,3),(7,3),(8,4), \ldots \}$ is such a sequence that converges to $1/2.$ Let nonempty $F_p \subset F$ be the set of all such $g_{ap}.$ Note that for the set $F_p,$ $p$ is fixed and $F_p$ contains each $g_{ap},$ as $a$ varies over $A,$ that satisfies the convergence requirement. Thus$,$ for $0\leq p \leq 1,$ $A$ is partitioned into subsets $A_p$ and a single set $A'$ such that each member of $A_p$ determines a $g_{ap} \in F_p.$ The elements of $A'$ are the members of $A$ that are not so characterized by such a $p.$ Let ${\cal A}$ denote this set of partitions.  \pars
Let $B = \{f\mid \forall n \forall m(((n \in \nat^{>0})\land (m \in \nat)\land (m \leq n)) \to ((f\colon ([1,n] \times \{n\}) \times \{m\} \to \{0,1\})\land (\forall j(((j \in \nat^{>0}) \land (1 \leq j \leq n)) \to (\sum_{j =1}^n f(((j,n),n),m) =m)))))\}.$\pars
\noindent The members of $B$ are determined$,$ but not uniquely$,$ by each $(n,m)$ such that $(n \in \nat^{>0})\land (m \in \nat)\land (m \leq n).$ Hence for each such $(n,m),$ let $f_{nm}\in B$ denote a member of $B$ that satisfies the conditions for a specific $(n,m).$ \par
For a given $p,$ by application of the axiom of choice$,$ with respect to ${\cal A},$ there is an $a \in A_p$ and a $g_{ap}$ with the properties discussed above. Also there is a  sequence $f_{na(n)}$ of partial sequences such that$,$ when $n >1,$ it follows that ($\dag$) 
$f_{na(n)}(j) = f_{(n-1)a(n-1)}(j)$ as $1\leq j \leq (n-1).$  Relative to the above example$,$ consider the following: 
$$f_{1a(1)}(1)  =0,$$
$$f_{2a(2)}(1)  =0,\ f_{2a(2)}(2)  =1,$$
$$f_{3a(3)}(1)  =0,\ f_{3a(3)}(2)  =1,\ f_{3a(3)}(3)  =0,$$
$$f_{4a(4)}(1)  =0,\ f_{4a(4)}(2)  =1,\ f_{4a(4)}(3)  =0,\ f_{4a(4)}(4) =1,$$
$$ f_{5a(5)}(1)  =0,\ f_{5a(5)}(2)  =1,\ f_{5a(5)}(3)  =0,\ f_{5a(5)}(4) =1,\ f_{5a(5)}(5) = 0, \cdots$$\par
It is obvious how this unique sequence of partial sequences is obtained from any $a \in A.$ For each $a \in A,$ let $B_a =\{f_{nm}\mid \forall n(n \in \nat^{>0} \to m = a(n))\}.$ Let $B_a^{\dag} \subset B_a$ such that each $f_{nm} \in B_a^{\dag}$ satisfies the partial sequence requirement ($\dag$). For each $n \in \nat^{>0},$ let $Pf_{na(n)}\in B_a^{\dag}$ denote the unique partial sequence of $n$ terms generated by
an $a$ and the ($\dag$) requirement. In  general$,$ as will be demonstrated below$,$ it is the $Pf_{na(n)}$ that yields the set of consequence operators as they are defined on $\rm L$ or ${\rm L}_1$ and as generated by expression [4]. Consider an additional map $M$ from the set $PF = \{Pf_{na(n)}\mid a\in A\}$ of these partial sequences into our  descriptive language $\rm L$ for the source $\rm G$  and an event (i.e. the 
$a_i$ of expression [4]). For each $n \in \nat^{>0},$ and $1 \leq j \leq n,$ if $Pf_{na(n)}(j) = 0,$ then $M(Pf_{na(n)}(j))= \r E'$ (i.e. ${\rm E' = E}$ does not occur); if $Pf_{na(n)}(j) = 1,$ then $M(Pf_{na(n)}(j))= \r E$ (i.e. E does occur)$,$ as 
$1\leq j \leq n,$ where the partial sequence $j = 1,\cdots, n$ models the intuitive concept of an event sequence (Herrmann$,$ 1994) in such a manner that each $\r E$ or $\r E'$ contains the appropriate Kleene ``tick'' symbols or natural number symbols that are an abbreviation for this tick notation. 
\parm 
Consider the set of consequence operators$,$ each defined on $\r L,$ $H = \{C(X,\{\r G\})\mid X \subset \r L \},$ where if  $\r G \in Y,$ then
$C(X, \{\r G\})(Y) = Y \cup X;$ if $\r G \notin Y,$ then $C(X, \{\r G\})(Y) = Y.$ Then for each $a \in A_p,$ $n \in \nat^{>0}$ and the respective $Pf_{na(n)},$ there exists the set of consequence operators $C_{ap}=
\{C(\{M(P_{na(n)}(j))\},\{\r G\})\mid 1\leq j \leq n\} \subset H.$
Note that from Herrmann (1987$,$ p. 5)$,$ $H$ is closed under the finite $\vee$ and the actual consequence operator is 
$C(\{M(P_{na(n)}(1))\} \cup\cdots \cup  \{M(P_{na(n)}(n))\}, \{\r G\}).$ Applying the realism relation to $C(\{M(P_{na(n)}(1))\} \cup \cdots \cup\{M(P_{na(n)}(n))\},\{\r G\})(\{\r G\})$ yields the actual labeled or identified event partial sequence $\{M(P_{na(n)}(1)), \ldots, M(P_{na(n)}(n))\}.$
\parm
\noindent \qquad Now imbed the above intuitive results into the superstructure 
${\cal M} = \langle {\cal N}, \in ,=\rangle$\break which is further imbedded into the nonstandard structure $\hyper {\cal M} = \langle \hyper {\cal N}, \in , = \rangle$ (Herrmann$,$ 1993). Let $p\in \real$ be such that $0\leq p\leq 1,$ where $p$ represents a theory predicted (i.e. a priori) probability that an event will occur. Applying the a choice function $C$ to $\cal A,$ there is some $a \in A_p$ such that $g_{ap} \to p.$ Hence let $\nu \in \hypernat$ be any infinite natural number. The hyperfinite sequence $\{ {a_1},\ldots,a_n,\ldots,\kern-.3em\hyper {a_\nu}\}$ exists and corresponds to $\{a_1,\ldots, a_n\}$ for any natural number $n \in \nat^{>0}.$ Also we know that $\st{\hyper a_\mu}= p$ for any infinite natural number $\mu.$ Thus there exists some 
internal hyperfinite $Pf_{\nu\kern-.3em\hyper {a(\nu)}} \in \hyper {PF}$ with the 
*-transferred properties mentioned above. Since $\Hyper {\b H}$ is closed under hyperfinite $\vee,$ there is a $P_p \in \Hyper {\b H}$ such that$,$ after application of the relation $\Hyper R,$ the result is the hyperfinite sequence $S=\{\kern-.3em\hyper {M}(P_{\nu\kern-.3em\hyper {a(\nu)}}(1)), \ldots,\kern-.3em\hyper {M}(P_{\nu\kern-.3em\hyper {a}(\nu)}(j)),\ldots, \kern-.3em\hyper {M}(P_{\nu\kern-.3em\hyper {a}(\nu)}(\nu))\}.$ Note that if $j\in \nat,$ then we have that $\Hyper {\b E} = \b E$ or $\Hyper {\b E'} = \b E'$ as the case may be. \par
An extended standard mapping that restricts $S$ to internal subsets would restrict $S$ to  $\{\kern-.3em\hyper {M}(P_{\nu\kern-.3em\hyper {a(\nu)}}(1)), \ldots,\hyper {M}(P_{\nu\kern-.3em\hyper {a}(\nu)}(j))\},$ whenever $j \in \nat^{>0}.$ Such a restriction map models the restriction of $S$ to the natural-world in accordance with the general interpretation given for internal or finite standard objects (Herrmann$,$ 1993$,$ p. 98)$,$ and the result either directly or indirectly affects the natural-world.  This completes the proof. \parm
In Theorem 2$,$ the force-like ultralogic $P_p$ is very unusual since$,$ although it satisfies the basic consequence operator properties on internal objects$,$ it does not seem to correspond to any ordinary form of human deduction for the image of $P_p$ on $\Hyper {\{{\b G}\}} = \{ \b G \}$ is an internal nonstandard object. Further$,$ the choice of a specific $a \in A_p$ is not an absolutely random choice since the set $A_p$ is a well-defined mathematical object. Indeed, the process is intuitively the same as the linguistic process of selecting a representative from the ``saying the same thing'' equivalences classes mentioned prior to the statement of Theorem 1. Consequently, the axiom of choice selection from a given $A_p$  can be considered as but an analogue to
this linguistics process. Further$,$ many predicted probabilities $p$ depend upon a particular state-of-affairs$,$ a particular scenario. An alteration in the scenario$,$ at any moment$,$ yields a possible new $p.$ 
There are NSP-world processes$,$ however$,$ that can transmit information identifying these alterations ``instantaneously$,$'' as perceived from the natural-world$,$ throughout our entire universe. This might influence the selection of the appropriate $p$ from  a different $A_p.$ For further details see section 11.2 of Herrmann (1993) or the Appendix of Herrmann (1999). In the laboratory setting$,$ although the probability $p$ may be accurately calculated$,$ the actual event partial sequence that yields the relative frequencies often seems to be very inaccurate as an approximator for $p.$ This is readily explained since there are $a \in A_p$ such that $g_{ap}$ converges ``slowly'' to $p.$ 
\parm
\centerline{\bf Distributions}\parm
Prior to considering the very important statistical notion of a frequency function and the distribution it generates$,$ there is need to consider a finite {\bf Cartesian product} consequence operator. Suppose that we have a finite set of consequence operators $\rm \{C_1, \ldots, C_m\},$ each defined upon its own language ${{\cal L}_k}$ and, at the least, one is axiomless. Define the operator $\rm \Pi C_m$ as follows: for any $\rm X \subset {\cal L}_1 \times \cdots \times {\cal L}_m$$,$ using the projections $\rm pr_k$$,$ consider the Cartesian product $\rm {pr_1(X)  \times \cdots \times pr_m(X)}$. Then 
$\rm \Pi C_m (X) = C_1(pr_1(X))\times \cdots \times C_m(pr_m(X))$ is a consequence operator on subsets of ${{\cal L}_1}\times \cdots \times {{\cal L}_m}.$ If each $\rm C_k$ is a finitary consequence operator$,$ then $\rm \Pi C_m$ is finitary. In all other cases$,$ $\rm \Pi C_m$ is a general consequence operator. All of these standard facts also hold within our nonstandard structure under *-transfer. \pars
A distribution's frequence function is always considered to be the probabilistic measure that determines the number of events that occur within a {\bf cell} or ``interval'' for  a specific decomposition of the events into various definable and disjoint  cells. There is a specific probability that a specific number of events will be contained in a specific cell and  each event must occur in one and only one cell and not occur in any other cell. \pars
For each distribution over a specific set of cells$,$ $I_k$$,$ there is a specific probability $p_k$ that an event will occur in  cell $I_k$. Assuming that the distribution does indeed depict physical behavior$,$ we will have a special collection of $g_{ap_k}$ sequences generated. For example$,$ assume that we have three cells and the three probabilities $p_1 = 1/4,\ p_2 = 1/2, \ p_3 = 1/4$ that events will occupy each of these cells. Assume that the number of events to occur is 6. Then the three partial sequence might appear as follows $$\cases{g_{ap_1}= \{(1,1),(2,1),(3,1),(4,2),(5,2), (6,2)\}&\cr
                    g_{ap_2}= \{(1,0),(2,1),(3,2),(4,2),(5,2),(6,3)\}&\cr
                    g_{ap_3}= \{(1,0),(2,0),(3,0),(4,0),(5,1),(6,1)\}&\cr}$$
Thus after six events have occurred$,$ 2 events are in the first cell$,$ 3 events are in the second cell$,$ and only 1 event is in the third cell. Of course$,$ as the number of events continues the first sequence will converge to 1/4$,$ the second to 1/2 and the third to 1/4. Obviously for any $n \geq 1, g_{ap_1}(n) + g_{ap_1}(n) + g_{ap_3}(n) = n.$ Clearly$,$ these required $g_{ap_i}$ properties can be generalized to any finite number $m$ of cells.
Relative to each factor of the Cartesian product set$,$ all of the standard aspects of Theorem 2 will hold. Further$,$ these intuitive results are embedded into the above superstructure and further embedded into our nonstandard structure. Hence$,$ assume that the languages $\rm L_k = L$ and that the standard consequence operators $C_k$ are all the same operators used to create a collection of product consequence operators where each factor 
is the $C_{ap_k}$ of Theorem 2. Under the nonstandard embedding$,$ we would have that for each factor$,$ there is a pure nonstandard consequence operator $P_{p_k} \in \Hyper {\b H_k}$. Finally$,$ consider the product consequence operator $\Pi P_{p_m}.$ For $\Hyper {(\{\b G_1\} \times \cdots \times \{\b G_m\})} = \{\b G_1\} \times \cdots \times \{\b G_m\},\ \b G_i = \b G$$,$ this nonstandard product consequence operator yields for any fixed event number $n$$,$ an ordered m-tuple$,$ where one and only one coordinate would have the statement $\b E$ (the event occurs) and all other coordinates the $\b E'$ (the event does not occur). It would be these m-tuples that guide the proper cell placement for each event and would satisfy the usual requirements of the distribution. Hence$,$ the patterns produced by a specific frequency function for a specific distribution may be rationally assumed to be the result of an application of an ultralogic.\parm

\centerline{\bf References}\parm
Cohen$,$ M. R. and E. Nagel. 1934. An introduction to logic and scientific method. Harcourt$,$ Brace and Co.$,$ New York.\parm
Feynman$,$ R. 1985.  QED The Strange Theory of Light and Matter. Princeton University Press. \parm
Herrmann$,$ R. A. 1987. Nonstandard Consequence Operators. {\it Kobe. J. Math.} 4:1-14. $<$http://www.arXiv.org/abs/math.GM/9911204$>$\parm
Herrmann$,$ R. A. 1993. Ultralogics and More. $<$http://www.arXiv.org/abs/math.GM/9903081$>$ or (math.HO/9903081) or (math/9903081)$,$ and /9903082\parm
Herrmann$,$ R. A. 1994. 
 Solutions to the ``General Grand Unification Problem$,$'' and the Questions ``How Did Our Universe Come Into Being?'' and ``Of What is Empty Space Composed?'' Presented before the MAA$,$ at Western Maryland College$,$ 12 Nov. $<$http://www.arXiv.org/abs/astro-ph/9903110$>$\parm
Herrmann$,$ R. A. 1999. The NSP-World and Action-at-a-Distance. in  Vol. 2$,$  {\it Instantaneous Action-at-a-Distance in Modern Physics: ``Pro'' and ``Contra''} ed. Chubykalo$,$ A.$,$  N. V. Pope and R. Smirnov-Rueda$,$ Nova Science Books and Journals$,$ New York\parm
Herrmann$,$ R. A. An ultimate hyperfinite ultralogic unification for allphysical theories$,$ $<$http://www.arXiv.org/abs/physics/0205073$>$ 
James and James. 1968. Mathematical Dictionary. D. Van Nostrand Co$,$ N. Y.\parm
Kleene$,$ S. 1967. Mathematical Logic. John Wiley\& Sons. New York.\parm
March$,$ A. and I. M Freeman. 1963. The New World of Physics. Vintage Books$,$ N. Y.\parm
Olkin$,$ I. et al. 1994. Probability Models and Applications. Macmillion College Publishing$,$ N.Y.\parm
Planck$,$ M. 1932.  The Mechanics of Deformable Bodies. Vol II. Introduction to Theoretical Physics. Macmillion$,$ N. Y.\parm
Preskill$,$ J. 1997. Quantum Information and Quantum Computing. Physics 229$,$ Advanced Mathematical Methods in Physics. Cal. Tech. \parm
W\'ojcicki$,$ R. 1970. Some remarks on the consequence operation in sentential logics. {\it Funda. Math.} 68:269-279.\parm
\end